\magnification 1200
\centerline {\bf Model of Antiferromagnetic Superconductivity}
\vskip 0.5cm
\centerline {{\bf by Geoffrey L. Sewell}\footnote*{e-mail address: 
g.l.sewell@qmul.ac.uk}}
\vskip 0.5cm
\centerline {\bf Department of Physics, Queen Mary University of 
London}
\vskip 0.2cm
 \centerline {\bf Mile End Road, London E1 4NS, UK} 
\vskip 1cm
\centerline {\bf Abstract}
\vskip 0.3cm
We present a simple model that supports coexistent superconductive and 
antiferromagnetic ordering. The model consists of a system of electrons on a simple 
cubic lattice that move by tunnel effect and interact via antiferromagnetic Ising spin 
couplings and short range repulsions: these include infinitely strong Hubbard forces 
that prevent double occupancy of any lattice site. Hence, under the filling condition 
of one electron per site and at sufficiently low temperature, the system is an 
antiferromagnetic Mott insulator. However, when holes are created by suitable 
doping, they are mobile charge carriers. We show that, at low concentration, their 
interactions induced by the above interelectronic ones lead to Schafroth pairing. 
Hence, under certain plausible but unproven assumptions, the model exhibits the off-
diagonal long range order that characterises superconductivity, while retaining the 
antiferromagnetic ordering. 
\vfill\eject
\centerline {\bf 1. Introduction}
\vskip 0.3cm
Since the discovery of superconducting phases in certain compounds 
[1], it has become clear that these generally come with rich arrays of 
neighbouring and even coexisting \lq normal\rq\ phases that include both 
magnetically ordered and insulating ones (cf. the books by Anderson [2] and Plakida 
[3] and references therein). There is now a huge literature devoted to quantum 
theories of these phase structures and this has provided a picture dominated by strong 
correlations due to intra-atomic Hubbard repulsive forces  (see Refs. [2]-[5] for 
example) . 
\vskip 0.2cm
However, in our view, it is still conceptually unclear how these correlations lead to 
superconductive ordering, and consequently we consider that there is a need for a 
model that reveals the mechanism behind this ordering in simple physical terms. The 
object of this article is to provide such a model.
\vskip 0.2cm
The model is a version of one devised by Richmond and the author [6], long before 
the discovery of superconducting compounds, for the purpose of providing  a 
treatment of magnetic ordering. It comprises a system, ${\Sigma}$, of $N$ electrons 
on a simple cubic periodicised lattice that move by tunnel effect and interact via (a) 
antiferromagnetic Ising couplings, (b) certain position dependent two-body 
repulsions of slightly longer range, and (c) infinitely strong Hubbard intra-atomic 
forces that prevent the occupation of each site by more than one electron and thus 
enforce an {\it extended exclusion principle}. Evidently, under the filling condition 
of one electron per site, the system is a Mott insulator. However, when the system is 
suitably doped, holes are created and these are mobile charge carriers. In the 
particular  case where there are just two holes, it is a simple matter to see that they 
bind together when the ratio of the Ising coupling parameter to the tunnelling one is 
sufficiently large: the repulsive interactions (b) do not affect this result as they are 
dominated by the Ising coupling at sufficiently short range. Thus, the prescribed 
interactions lead to a {\it Schafroth pairing} [7], the bound pair constituting a boson. 
Since this arises from position dependent interactions, it differs from the Cooper 
pairing [8] of metallic superconductors, which stem from momentum dependent 
ones: see also the discussion by Leggett [9] of these two kinds of pairing. Further, in 
the physically interesting case where the number of holes is macroscopic and their 
density is low, the pairing prevails and the repulsive interactions (b), if sufficiently 
strong, prevent the formation of larger bound hole clusters. Consequently the model 
reduces to a dilute gas of charged bosons and, under certain natural but unproven 
assumptions, it executes a Bose-Einstein (BE) condensation. More precisely, it 
exhibits off-diagonal long range order (ODLRO), which is the generalised form of 
BE condensation, applicable even to systems of interacting bosons [10, 11]. As 
shown in Refs. [12]-[14], ODLRO implies the characteristic electromagnetic 
properties of superconductors with ordering represented by a macroscopic wave 
function. 
\vskip 0.2cm
We present our description and treatment of the system, ${\Sigma}$, as follows. In 
Section 2 we formulate the model of this system in precise mathematical terms. Here 
the extended exclusion principle, noted above, leads to a key modification of the 
canonical commutation relations [6]. Further, under the filling condition of one 
electron per site, it forbids any tunnelling and therefore the system reduces to an 
antiferromagnetic Mott insulator. However, the introduction of holes, due to suitable 
doping, provides the model with charge carriers. 
\vskip 0.2cm
In Section 3, we formulate the system, $S^{(n)}$, obtained by the introduction of an 
arbitrary number, $n$, of holes into the ground state of ${\Sigma}$. Here the  key 
Proposition 3.1, whose proof is left to the Appendix, serves to pass from the second 
quantisation picture, {\it as modified by the extended exclusion principle}, to a 
simple first quantisation picture of the $n$ hole system. In Section 4 we specialise to 
the situation where there are just two holes. There we show that, under conditions 
wherein the Ising coupling parameter is sufficiently large by comparison with the 
tunnelling one, the holes bind together. As noted above, this corresponds to 
Schafroth pairing [7]. On the other hand, we note that, in the situation where more 
than two holes are introduced, the repulsive interactions render the formation of 
larger spatial clusters of them energetically unfavourable, in that their potential 
energies are not minimal.
\vskip 0.2cm  
In Section 5 we invoke the observations of the previous Section in our treatment of 
the model $S^{(2n)}$ for $n=O(N)$ and the density $2n/N<<1$. Specifically, we 
assume that the pairing mechanism still prevails and that spatial clusters of more than 
two holes do not occur. These assumptions essentially constitute an Ansatz, 
according to which $S^{(2n)}$ reduces to a system of $n$ bosonic \lq atoms\rq , 
each atom being a Schafroth pair. 
\vskip 0.2cm
In Section 6, we consider the question of whether the model exhibits BE 
condensation, as represented by ODLRO, in a limit where $N$ becomes infinite and 
the hole density ${\eta} \  (=2n/N)$ is fixed. There  we note that, under conditions of 
strong repulsive interactions and low density ${\eta}$, together with a supplementary 
scaling assumption, this model simulates that of Lieb et al [15], which has been 
proved to undergo BE condensation. On this basis, we assume that the ground state 
of the model satisfies the ODLRO condition and hence that it is superconductive. We 
further assume that the magnetic ordering of ${\Sigma}$  prevails at sufficiently low 
hole density. Hence we conclude that the system is an antiferromagnetic 
superconductor in its ground state. This result is in line with that obtained on purely 
thermodynamic grounds by Kivelson et al [16]. \vskip 0.2cm
We conclude in Section 7 with a brief discussion of open problems
\vskip 0.5cm
\centerline {\bf 2. The Model}
\vskip 0.3cm
{\bf 2.1. The Lattice $X$.} We assume that the model is a system, 
${\Sigma}$, of  electrons that live on the periodicised $d$-dimensional simple cubic
lattice $X$ of side $L$, i.e, the discrete space $\bigl(Z \ ({\rm mod}L)\bigr)^{d}$, 
whose points $x$ are $d$-tuples $(x^{1},. \ .,x^{d})$, where each $x^{j}$ 
runs through the set of integers $(1,2,. \ .,L)$ whose ends are identified 
with one another. Thus the number, $N$, of sites of $X$ is $L^{d}$. We 
assume that $L$ is an even integer.  In anticipation of an antiferromagnetic 
configuration of the electrons, we resolve $X$ into interlocking subspaces $X^{+}$ 
and  $X^{-}$, the former consisting of the points $x=(x^{1},. \ .,x^{d})$ for which 
${\sum}_{j=1}^{d}x^{j}$ is even and the latter for which that sum is odd. We 
denote by ${\cal U}$ and ${\cal V}$ the sets of nearest neighbours of the origin, 
$O$, in $X$ and $X^{+}$, respectively.
\vskip 0.3cm
{\bf 2.2. State Space and Algebraic Structure.} In general, the pure states 
of a many-electron system are represented by the normalised vectors in a Fock 
Hilbert space ${\cal F}$. However, in view of the above discussed 
extended exclusion principle, the states of the present model are restricted 
to the subspace ${\cal H}$ of ${\cal F}$ for which no more than one 
electron can be situated at any site. We formulate the spaces 
${\cal F}$ and ${\cal H}$ as follows.
\vskip 0.2cm
The Fock space ${\cal F}$ may be defined in terms of a vacuum vector ${\Omega}$ 
and bounded operators $c^{\star}(x,{\lambda})$ and $c(x,{\lambda})$ governing 
the creation and annihilation of an electron of spin ${\lambda} \ (={\pm}1)$ at the 
site $x$. The defining properties of ${\Omega}$ and $c(x,{\lambda})$ are that
\vskip 0.2cm\noindent
(i) ${\Omega}$ is annihilated by each of the operators $c(x,{\lambda})$;
\vskip 0.2cm\noindent
(ii) ${\Omega}$ is cyclic with respect to the adjoints 
$c^{\star}(x,{\lambda})$ of 
these operators, i.e. the vectors obtained by action on ${\Omega}$ of the 
polynomials in these adjoints form a dense subset of ${\cal F}$; and 
\vskip 0.2cm\noindent
(iii) the operators $c$ and $c^{\star}$ satisfy the canonical 
anticommutation relations
$$\bigl[c(x,{\lambda}),c^{\star}(x^{\prime},{\lambda}^{\prime})]_{+}
={\delta}(x,x^{\prime})
{\delta}({\lambda},{\lambda}^{\prime})I_{\cal F} \ {\rm and}
\bigl[c(x,{\lambda}),c(x^{\prime},{\lambda}^{\prime})\bigr]_{+}=0 \ 
{\forall} \ x,x^{\prime}{\in}X; \ {\lambda}, \ 
{\lambda}^{\prime}={\pm}1,\eqno(2.1)$$
where the ${\delta}$ is that of Kroneker and $I_{\cal F}$ is the identity 
operator in ${\cal F}$. The number of electrons of spin ${\lambda}$ at the site $x$ 
is represented by the operator 
$${\nu}_{\cal F}(x,{\lambda})=c^{\star}(x,{\lambda})c(x,{\lambda}).\eqno(2.2)$$ 
The Hilbert space ${\cal H}$, which harbours the states of the system, subject to the 
extended exclusion condition, is the subspace $P{\cal F}$ of ${\cal F}$, where $P$ 
is the Gutzwiller projector  [17] given by the following formula (see also Ref. [6]).
$$P={\Pi}_{x{\in}X}\bigl(I_{\cal F}-{\nu}_{\cal F}(x,1)
{\nu}_{\cal F}(x,-1)\bigr).\eqno(2.3)$$
Correspondingly, under the same condition, the creation and annihilation operators 
for an electron of spin ${\lambda}$ at the site $x$ are
$$a^{\star}(x,{\lambda})=Pc^{\star}(x,{\lambda})P \ {\rm and} \ 
a(x,{\lambda})=Pc(x,{\lambda})P,\eqno(2.4)$$
respectively. Thus, under this condition, the number of electrons of spin 
${\lambda}$ at the site $x$ is given by the operator
$${\nu}(x,{\lambda})=a^{\star}(x,{\lambda})a(x,{\lambda}),\eqno(2.5)$$
which, in view of Eqs. (2.1)-(2.5), is equal to $P{\nu}_{\cal F}(x,{\lambda})P$. The 
total number of particles, ${\nu}(x)$, and the spin, ${\sigma}(x)$, at $x$ are 
therefore given by the formulae
$${\nu}(x)={\sum}_{{\lambda}={\pm}1}{\nu}(x,{\lambda})\eqno(2.6)$$
and
$${\sigma}(x)={\sum}_{{\lambda}={\pm}1}{\lambda}{\nu}(x,{\lambda}).
\eqno(2.7)$$
Further, as the projector $P$ is the identity operator, $I$, of the space 
${\cal H}$, it follows from Eqs. (2.1)-(2.7) that the operators $a, \ 
a^{\star}, \ {\nu}$ and ${\sigma}$ satisfy the following algebraic relations, which 
were established in Ref.  [6].
$$\bigl[a(x,{\lambda}),a(x^{\prime},{\lambda}^{\prime})\bigr]_{+}=0,
\eqno(2.8)$$
$$\bigl[a(x,{\lambda}),a^{\star}(x^{\prime},{\lambda}^{\prime})
\bigr]_{+}=\bigl(I-{\nu}(x,-{\lambda})\bigr){\delta}(x,x^{\prime})
{\delta}({\lambda},{\lambda}^{\prime})+$$
$$a^{\star}(x,{\lambda})a(x,-{\lambda}){\delta}(x,x^{\prime})
{\delta}({\lambda},-{\lambda}^{\prime}),\eqno(2.9)$$
$$a^{\star}(x,{\lambda})a^{\star}(x,{\lambda}^{\prime})=0 \ {\forall} \ x{\in}X, \ 
{\lambda},{\lambda}^{\prime}={\pm}1.\eqno(2.10)$$
and
$$a(x,{\lambda}){\Omega}=0.\eqno(2.11)$$
\vskip 0.3cm
{\bf 2.3. Hamiltonian of the Model.} We assume that this is the operator, $H$, in 
${\cal H}$ that takes the form
$$H=H^{T}+H^{J}+H^{K},\eqno(2.12)$$
the three contributions representing tunnelling, Ising spin couplings and 
particle repulsions, respectively. These are defined so that the Ising interactions 
couple particles at nearest neighbouring sites, the repulsive interactions couple next 
nearest neighbouring ones, and the tunnelling processes are spin conserving 
transitions between the nearest sites that harbour particles of the same spin in the 
Ising antiferromagnetic ground state.  Thus the Hamiltonian components are assumed 
to take the following forms.
$$H^{T}=T{\sum}_{x{\in}X;v{\in}{\cal V}}
\bigl(a^{\star}(x,{\lambda})a(x+v,{\lambda})
+h.c.\bigr){\equiv}-T{\sum}_{x{\in}X,v{\in}{\cal V}}
\bigl(a(x+v,{\lambda})a^{\star}(x,{\lambda})+h.c.\bigr),\eqno(2.13)$$
$$H^{J}=J{\sum}_{x{\in}X; \ u{\in}
{\cal U}}{\sigma}(x){\sigma}(x+u),\eqno(2.14)$$
; and
$$H^{K}=K{\sum}_{x{\in}X;u,u^{\prime}({\neq}-u){\in}{\cal U}}
{\nu}(x){\nu}(x+u+u^{\prime}),\eqno(2.15)$$
where $T, J$ and $K$ are all positive and ${\cal U}$ and ${\cal V}$ are the subsets 
of $X$ defined in Section 2.1
\vskip 0.3cm	 
{\bf 2.4. The Filling Condition, Mott Insulation and Antiferromagnetic Order}  
Under the filling condition, where the total number of electrons is equal to $N$, the 
number of lattice sites, the extended exclusion principle signifies that each site of 
$X$ is occupied by precisely one electron. Hence this condition signifies that the 
state space of ${\Sigma}$ is reduced to the subspace of  ${\cal H}$ for which the 
following vectors ${\Psi}({\lambda})$, which correspond to spin configurations  
${\lbrace}{\lambda}(x){\vert}x{\in}X{\rbrace}$ form an orthogonal basis. 
$${\Psi}({\lambda})=
\bigl[{\Pi}_{x{\in}X}a^{\star}\bigl(x,{\lambda}(x)\bigr)\bigr]{\Omega}.\eqno(2.16)
$$
It follows from this formula and Eqs. (2.6)-(2.10) that 
$$a^{\star}\bigl(x,{\lambda}^{\prime}(x)){\Psi}({\lambda})=0,\eqno(2.17)$$
$${\nu}(x){\Psi}({\lambda})=
{\Psi}({\lambda}) \ {\forall} \ x{\in}X.\eqno(2.18)$$ 
and
$${\sigma}(x){\Psi}({\lambda})={\lambda}(x){\Psi}({\lambda}).\eqno(2.19)$$
\vskip 0.2cm
It now follows from Eqs. (2.10), (2.13) and (2.16) that
$$H^{T}{\Psi}({\lambda})=0,\eqno(2.20)$$  
which signifies that the extended exclusion prevents any tunnelling and 
thus produces Mott insulation. Further, by Eqs. (2.15) and (2.18), $H^{K}$ reduces 
to a constant, namely $NKd$, on ${\Psi}({\lambda})$. Therefore, in view of  Eqs. 
(2.12) and (2.20), the effective Hamiltonian reduces to $H^{J}$, and 
consequently the equilibrium state of the system is antiferromagnetically ordered 
below the critical Ising temperature. Hence the ground states of the model are just 
those of the Ising system, as governed by the Hamiltonian $H^{J}$. There are just 
two of these, which we denote by ${\Phi}$ and ${\Phi}^{\prime}$ and which, by 
Eqs. (2.14) and (2.19), are simultaneous eigenvectors of the ${\sigma}(x)$\rq s with 
corresponding eigenvalues ${\lbrace}s(x)={\pm}1{\forall}x{\in}X^{\pm}{\rbrace}$ 
and ${\lbrace}s(x)={\mp}1{\forall}x{\in}X^{\pm}{\rbrace}$, respectively. Our 
treatment of the model will be centred on operations applied to the state vector 
${\Phi}$: the corresponding treatment based on ${\Phi}^{\prime}$ and its 
modifications could be carried out analogously. 
\vskip 0.2cm
It follows from the above specifications that
$${\sigma}(x){\Phi}=s(x){\Phi} \ {\forall} \ x{\in}X\eqno(2.21)$$
where
$$s(x)=1 \ {\forall} \ x{\in}X^{+} \ {\rm and} \ -1 \ 
{\forall} \ x{\in}X^{-},\eqno(2.22)$$
and
$${\Phi}=\bigl[{\Pi}_{x{\in}X}a^{\star}\bigl(x,s(x)\bigr)\bigr]{\Omega}.
\eqno(2.23)$$
We remark here that it may be assumed, without loss of generality, that 
$$H{\Phi}=0,\eqno(2.24)$$
since this equation is validated by adding the constant $-NKd$  to the r.h.s. of Eq. 
(2.12). 
\vskip 0.2cm
Suppose now that the spin configuration ${\lbrace}s(x){\vert}x{\in}X{\rbrace}$, 
corresponding to the ground state ${\Phi}$, is modified by reversal of the spins in a 
subset ${\Delta}$ of $X$. The resultant state of ${\Sigma}$ then becomes
$${\Phi}_{\Delta}=
\bigl[{\Pi}_{x{\in}X{\backslash}{\Delta}}a^{\star}\bigl(x,s(x)\bigr)\bigr]
\bigl[{\Pi}_{x{\in}{\Delta}}a^{\star}\bigl(x,-
s(x)\bigr)\bigr]{\Omega}.\eqno(2.25)$$
It then follows that, corresponding to Eq. (2.24) for the ground state ${\Phi}$, 
$$H{\Phi}_{\Delta}=E({\Delta}){\Phi}_{\Delta},\eqno(2.26)$$
where $E({\Delta})$ is the energy required to reverse the spins in ${\Delta}$ when 
the system is in the ground state ${\Phi}$. 
\vskip 0.5cm  
\centerline {\bf 3. Holes in the Antiferromagnetically Ordered Ground State }
\vskip 0.3cm
We now consider the modification of the state ${\Phi}$ by the introduction of $n$ 
holes, denoting the resultant system by $S^{(n)}$. We represent the creation of a 
hole at the site $x$ by the operator ${\alpha}(x)$, defined by the equation
$${\alpha}(x)=a\bigl(x,s(x)\bigr) \ {\forall} \ x{\in}X.\eqno(3.1)$$
The algebraic properties of the operators ${\alpha}(x)$ then follow from this formula 
and Eqs. (2.8)-(2.11). 
\vskip 0.2cm
{\bf 3.1. The System $S^{(n)}$.} We define $S^{(n)}$ to be the system of $n$ holes 
in the state ${\Phi}$ and represent its pure states by the linear combinations of the 
actions on ${\Phi}$ of the monomials of order $n$  in the ${\alpha}(x)$\rq s. Thus, 
in view of the anticommutation relations of the latter, these state vectors  take the 
form
$$C_{n}(f):=(n!)^{-1/2}{\sum}_{x_{1},. \ .,x_{n}{\in}X}f(x_{1},. \ 
.,x_{n}){\alpha}(x_{1}).\ .{\alpha}(x_{n}){\Phi},\eqno(3.2)$$
where $f$ is an element of the Hilbert space, ${\cal H}^{(n)}$, of antisymmetric 
function on $X^{n}$ with $l^{2}(X^{n})$ inner product. It follows from these 
specifications that 
$${\langle}C_{n}(f),C_{n}(g){\rangle}_{\cal H}=
{\langle}f,g{\rangle}_{{\cal H}^{(n)}} \ {\forall} \ f,g{\in}{\cal H}^{(n)}.
\eqno(3.3)$$
Thus, $C_{n}$ is an isomorphism of ${\cal H}^{(n)}$ into ${\cal H}$ and 
therefore the pure states of $S^{(n)}$ are faithfully  represented by the vectors in 
${\cal H}^{(n)}$. Correspondingly, the observables of $S^{(n)}$ are represented by 
the self-adjoint operators in this space. The following Proposition, which will be 
proved in the Appendix, provides the correspondence between first and second 
quantisation pictures of $S^{(n)}$ {\it under the condition of the  extended exclusion 
principle.} 
\vskip 0.3cm
{\bf Proposition 3.1.} {\it It follows from the above definitions that $H$ 
induces a self-adjoint operator $H^{(n)}$ in ${\cal H}^{(n)}$, which we take to be 
the Hamiltonian of $S^{(n)}$, according to the following formula.
$$HC_{n}(f)=C_{n}(H^{(n)}f) \ {\forall} \ 
f{\in}{\cal H}^{(n)}.\eqno(3.4)$$
where 
$$H^{(n)}=-T{\sum}_{j=1}^{n}
{\Delta}_{{\cal V}j}+{\sum}_{j,k=1}^{n}V(x_{j}-x_{k}), \eqno(3.5)$$
$$V(x_{j}-x_{k})=-J{\sum}_{u{\in}{\cal U}}
\bigl({\delta}(x_{j},x_{k}+u)+{\delta}(x_{j},x_{k}-u)\bigr)+$$
$$K{\sum}_{u,u^{\prime}({\neq}-u){\in}{\cal U}}
\bigl({\delta}(x_{j},x_{k}+u+u^{\prime})+{\delta}(x_{j},x_{k}-u-
u^{\prime})\bigr).\eqno(3.6)$$
and ${\Delta}_{{\cal V}j}$ is the versions for the site $x_{j}$ of the discretised 
Laplacian  ${\Delta}_{\cal V}$ defined by the formulae
$${\Delta}_{{\cal V}}f(x)={\sum}_{v{\in}{\cal V}}[f(x+v)+
f(x-v)-2f(x)] .\eqno(3.7)$$
Evidently, the first and second sums in Eq. (3.5) correspond to the kinetic and 
potential energy, respectively, of $S^{(n)}$.}
\vskip 0.3cm
{\bf Discussion.} This Proposition signifies that ${\cal H}^{(n)}$ is the state space 
and $H^{(n)}$ is the Hamiltonian for the system, $S^{(n)}$, of $n$ holes in the state 
${\Phi}$. We take this to be the model of the $n$ hole system at sufficiently low 
values of the ambient temperature and hole density $n/N$, even though this model 
differs from that of $n$ holes in the full system ${\Sigma}$, since that can also carry 
holes in the modifications, ${\Phi}_{\Delta}$, of ${\Phi}$ by reversals of finite 
numbers of spins, as defined by Eq. (2.25).  In fact our neglect of these modifications 
is based on the assumption that they do not significantly affect the qualitative 
features of the equilibrium states of the model at sufficiently low temperature and 
hole density $n/N$. The heuristic basis of this assumption is the following. First, in 
the absence of holes, the energy cost, given by Eq. (2.26), of the spin reversals would 
freeze them out at sufficiently low temperatures. The creation of holes, however, 
would change this situation since their interaction with the spin reversals would 
render this freezing incomplete. Nevertheless, we anticipate that this interaction is 
effectively very weak at sufficiently low hole density and consequently that the 
equilibrium properties of $S^{(n)}$ should simulate those of $n$ holes in the full 
system ${\Sigma}$ at sufficiently low temperatures and hole density. 
\vskip 0.5cm
 \centerline {\bf 4. Binding of Hole pairs.}
\vskip 0.3cm
Suppose now that just a pair of holes is introduced into the system when 
that is in its ground state. In order to obtain a condition for their binding, 
we first note that, by Eqs. (3.5)-(3.7), the Hamiltonian $H^{(1)}$ for a 
single hole is $-T{\Delta}_{\cal V}$ and, by Eq. (3.6), its spectrum is $[0,T]$. 
Hence, the condition for the existence of bound states of the pair of holes 
is that the spectrum of $H^{(2)}$ is at least partly in ${\bf R}_{-}$ or, 
equivalently, that there exists an element $h^{(2)}$ of ${\cal H}^{(2)}$ 
such that $(h^{(2)},H^{(2)}h^{(2)})<0$. The following proposition 
provides just such a condition.
\vskip 0.3cm
{\bf Proposition 4.1.} {\it A sufficient condition for the binding of the hole 
pair is that
$$J> {\gamma}T,\eqno(4.1)$$
where ${\gamma}$ is the number of elements of  ${\cal V}$, i.e. $2d(d-1)$.}
\vskip 0.3cm
{\bf Proof.} Choose 
$$h^{(2)}(x_{1},x_{2})=2^{-1/2}\bigl[{\delta}(x_{1},a){\delta}(x_{2},b)-
{\delta}(x_{1},b){\delta}(x_{2},a)\bigr],\eqno(4.2)$$
where $a$ and $b$ are nearest neighbouring points of $X$. Then by 
Eqs. (3.5)-(3.7) and (4.2), 
$$(h^{(2)},H^{(2)}h^{(2)}){\leq}2({\gamma}T-J),\eqno(4.3)$$
Hence, Eq. (4.1) is a sufficient condition for a bound state. 
\vskip 0.3cm 
{\bf 4.1. Internal and External Descriptions.} We now define $y$ and $z$ to be the 
relative displacement and the centre of the coordinate pair $(x_{1},x_{2})$ 
respectively, i.e.
$$ y=x_{1}-x_{2} \ {\rm and} \ z={1\over 2}(x_{1}+x_{2}).\eqno(4.4)$$
Thus $y$ and $z$ run over the spaces $X$ and   
$X/2:={\lbrace}({\xi}/2){\vert}{\xi}{\in}X{\rbrace}$, respectively. We refer to $y$ 
and $z$ as the internal and external coordinates, respectively, of the pair. Defining 
${\tilde X}$ to be the subspace  of $X{\times}(X/2)$ comprising the pairs $(y,z)$ 
for which $(z{\pm}y/2){\in}X$, we re-express the wave functions $f(x_{1},x_{2})$ 
as vectors in the Hilbert space ${\hat {\cal H}}^{(2)}:=
{\lbrace}{\hat f}{\in}l^{2}({\tilde X}){\vert}{\hat f}(-y,z)=
-{\hat f}(y,z){\rbrace}$, according to the formula
$${\hat f}(y,z):=f(x_{1},x_{2})=f(z+y/2,z-y/2).\eqno(4.5)$$
We define the representation spaces for the internal and external observables to be 
the Hilbert spaces ${\hat {\cal H}}_{\rm int}:=
{\lbrace}g{\in}l^{2}(X){\vert}g(x)=-g(-x){\rbrace}$ and 
${\hat {\cal H}}_{\rm ext}:=l^{2}(X/2)$, respectively. It follows from these 
specifications that ${\hat {\cal H}}^{(2)}=
{\hat {\cal H}}_{\rm int}{\otimes}{\hat {\cal H}}_{\rm ext}$. We denote by $U$ 
the unitary tansformatiom $f{\rightarrow}{\hat f}$ of ${\cal H}^{(2)}$ onto ${\hat 
{\cal H}}^{(2)}$ given by Eq. (4.5).
\vskip 0.2cm
We denote by $W$ the density matrix in  ${\cal H}^{(2)}$ representing the zero 
temperature state of $S^{(2)}$.  Then this state may also be represented in 
${\hat {\cal H}}^{(2)}$ by its unitary transform 
$${\hat W}=UWU^{\star}.\eqno(4.6)$$
We denote by ${\hat W}_{\rm int}$ the density matrix in 
${\hat {\cal H}}_{\rm int}$ representing the interior state of the pair, as defined by 
the formula 
$${\hat W}_{\rm int}={\rm Tr}_{\rm ext}({\hat W}),\eqno(4.7)$$
where ${\rm Tr}_{\rm ext}$ signifies partial trace over ${\hat {\cal H}}_{\rm ext}$. 
\vskip 0.3cm
{\bf 4.2. Hamiltonian Operators.} In view of Eq. (4.5), the Hamiltonian operator for 
the pair is given by the operator ${\hat H}^{(2)}$ in ${\hat {\cal H}}^{(2)}$ defined 
by the formula
$$[{\hat H}^{(2)}{\hat f}](y,z)=[H^{(2)}f](z+y/2,z-y/2).\eqno(4.8)$$
It follows from Eqs. (3.5)-(3.7), (4.5) and (4.8) that 
$$[{\hat H}^{(2)}{\hat f}](y,z)=$$
$$-T{\sum}_{v{\in}{\cal V}}\bigl({\hat f}(y+v,z+v/2)+{\hat f}(y-v,z-y/2)+
{\hat f}(y-v,z+v/2)+{\hat f}(y+v,z-v/2)-4{\hat f}(y,z)\bigr)+$$
$$V(y){\hat f}(y,z).\eqno(4.9)$$
Consequently, defining displacement operators $D_{\rm int}(a)$ and 
$D_{\rm ext}(b)$ in ${\hat {\cal H}}^{(2)}$ by the equations
$$[D_{\rm int}(a){\hat f}](y,z)={\hat f}(y-a,z) \ {\rm and} \ [D_{\rm ext}(b)
{\hat f}](y,z)={\hat f}(y,z-b),\eqno(4.10)$$
$${\hat H}^{(2)}=-T{\sum}_{v{\in}{\cal V}}
\bigl[G_{\rm int}(v)G_{\rm ext}(v)-4\bigr]+V(y),\eqno(4.11)$$
where
$$G_{\rm int}(v)=D_{\rm int}(v)+D_{\rm int}(-v)\eqno(4.12)$$
and
$$G_{\rm ext}(v)= D_{\rm ext}(v/2)+D_{\rm ext}(-v/2).\eqno(4.13)$$
\vskip 0.2cm 
We define the exterior Hamiltonian for the pair to be the partial mean of 
${\hat H}^{(2)}$ w.r.t. its internal variables, as governed by the density matrix  
${\hat W}_{\rm int}$, i.e.
$${\hat H}_{\rm ext}={\rm Tr}_{\rm int}({\hat W}_{\rm int}
{\hat H}^{(2)}),\eqno(4.14)$$ 
where  ${\rm Tr}_{\rm int}$ is the partial trace over 
${\hat {\cal H}}_{\rm int}^{(2)}$. Hence, by Eq. (4.11) with the constant term $-4$ 
in the square brackets excluded,
$${\hat H}_{\rm ext}=-T{\sum}_{v{\in}{\cal V}}
{\rm Tr}_{\rm int}\bigl({\hat W}_{\rm int}G_{\rm int}(v)\bigr)
G_{\rm ext}(v) +{\rm Tr}_{\rm int}\bigl({\hat W}_{\rm int}V(y)\bigr) 
\eqno(4.15)$$
We now note that the Hamiltonian of the model ${\Sigma}$ is invariant under 
interchanges of the principal axes of the lattice $X$, and we assume that the density 
matrix $W$ retains this symmetry. By Eqs. (4.7) and (4.12), this implies that the 
values of the partial trace ${\rm Tr}_{\rm int}\bigl({\hat W}_{\rm int}
G_{\rm int}(v)\bigr)$ is independent of $v$. We denote this quantity by $g$, a 
constant. Further, since Equ. (4.7) implies that 
${\rm Tr}_{\rm int}\bigl({\hat W}_{\rm int}V(y)\bigr)$ is also a constant, it follows 
from these observations that Eq. (4.15), as modified by the addition of certain 
harmless constants, takes the following explicit form.
$${\hat H}_{\rm ext}=-{\hat T}{\Delta}_{\cal V},\eqno(4.16)$$
where ${\Delta}_{\cal V}$ are the discretised Laplacians defined by Eq.. (3.7) and 
$${\hat T}=gT.\eqno(4.17)$$
Thus the Hamiltonian $H_{\rm ext}$ is that of a particle that moves freely on the 
space $X/2$.. 
\vskip 0.3cm
{\bf 4.2. Discount of Larger Clusters of Holes.} The proof of Prop. 4.1 exploited the 
fact that the potential energy of the pair, as given by the second sum in Eq. (3.5), is 
minimised when the two holes are nearest neighbours. One may reasonably ask 
whether larger stable clusters of holes may be formed in the case when there are 
more than two of them. Although we do not have a definite answer to this question, 
we shall now show that clusters of more than two holes cannot minimise the potential 
energy if the repulsion parameter $K$ is sufficiently large; and consequently that, in 
this case, the mechanism of bound pairing does not extend to larger clusters of holes. 
Our argument runs as follows. We define a cluster to be a set of holes that occupies a 
subset, $C$, of $X$ such that
\vskip 0.2cm\noindent
(i) each point of $C$ has a nearest neighbour in that set, and 
\vskip 0.2cm\noindent
(ii) at least one element of each pair of nearest neighbours in $C$ has another nearest 
neighbour in $C$. 
\vskip 0.2cm\noindent
Suppose that now $a$ and $b$ are a nearest neighbouring pair within a cluster $C$. 
Then at least one of these points, say $b$, has another nearest neighbour, $c$, in 
$C$. Hence, by Eqs. (3.6), the particle at  $a$ is coupled to that at $b$ with energy 
$-2J$ and to that at $c$ with energy $2K$. Further, the number of nearest neighbours 
of $a$ in $C$ cannot exceed $2d$, while the number of its next nearest neighbours is 
at least $1$. Hence the increment in the potential energy of the system due to the 
transfer of the hole to another region is no greater than $2dJ-K$. Thus the potential 
energy is not minimised by clusters of more than two holes if $K>2dJ$. Of course, 
this does not rule out the possibility that some quantum mechanism might lead to 
cluster-like correlations, but we shall ignore that possibility in the treatment that 
follows. This procedure amounts to an Ansatz.
\vskip 0.5cm 
\centerline {\bf 5. Bosonisation} 
\vskip 0.3cm 
We now pass to a treatment of the $2n$ hole system $S^{(2n)}$, with $n=O(N)$, 
and the density ${\eta} \ (=2n/N) <<1$. In accordance with the observation of the 
previous Section, we assume that the pair binding mechanism, established for the 
case where there are just two holes, still prevails for $S^{(2n)}$, and that the 
repulsive interactions prevent the formation of bound clusters of more than two 
holes. 
\vskip 0.2cm
Thus we base our treatment on the Ansatz that the holes of $S^{(2n)}$ combine into 
$n$ pairs. Evidently these are bosons, as its wave functions 
$f(x_{1},x_{2},. \ .,x_{2n})$ change sign under interchanges 
$x_{j}{\rightleftharpoons}x_{k}$ of single positions and therefore remain invariant 
under interchanges $(x_{j},x_{j+1}){\rightleftharpoons}(x_{k},x_{k+1})$ of  pairs. 
Hence the pairs behave as bosonic atoms, and we shall refer to them as such. We 
formulate the model of  these $n$ atoms by a natural generalisation of the derivation, 
in Section 4.1, of the exterior Hamiltonian of just one of these.
\vskip 0.2cm
To this end, we resolve the set $(x_{1},x_{2},.  \ .,x_{2n})$ of hole coordinates into 
pairs whose relative displacements are $y_{1},. \ .,y_{n}$ and whose centres are 
$z_{1},. \ .,z_{n}$, respectively.Thus the coordinates of the $n$ pairs of holes 
comprising $S^{(2n)}$ are 
$$(z_{1}+y_{1}/2,z_{1}-y_{1}/2), \ (z_{2}+y_{2}/2,z_{2}-y_{2}/2),. \ .,,
(z_{n}+y_{n}/2,z_{n}-y_{n}/2).$$
\vskip 0.2cm
To formulate the internal and external features of $S^{(2n)}$, we define the Hilbert 
spaces ${\hat {\cal H}}_{\rm int}^{(n)},  \ {\hat {\cal H}}_{\rm ext}^{(n)}$ and 
${\hat {\cal H}}^{(2n)}$ to be the natural $2n$ hole generalisations of the spaces 
${\hat {\cal H}}_{\rm int}, \  {\hat {\cal H}}_{\rm ext}$ and ${\hat {\cal 
H}}^{(2)}$, respectively, on which the description of pairs was based in Section 4. 
Thus we define ${\hat {\cal H}}^{(2n)}$ to be 
${\hat {\cal H}}_{\rm int}^{(n)}{\otimes}{\hat {\cal H}}_{\rm ext}^{(n)}$, where 
${\hat {\cal H}}_{\rm ext}^{(n)}$ and ${\hat {\cal H}}_{\rm int}^{(n)}$ are the 
bosonic and fermionic spaces given by the symmetric (resp. antisymmetric) 
elements, of  $l^{2}\bigl((X/2)^{n}\bigr)$  (resp. $l^{2}(X^{n})$), with the added 
restriction that the latter elements change sign under reversal of each component 
$y_{j}$ of the points $(y_{1},. \ .,y_{n})$ of the  underlying space $X^{n}$.
\vskip 0.2cm 
We take the internal state of each of these atoms to be given by the canonical copy of 
that of a single pair, as represented by the density matrix, ${\hat W}_{\rm int}$, 
formulated in Section 4.1; and we assume that, in view of the diluteness of the 
atomic system, these states may be treated as statistically independent. Thus the 
density matrix in ${\hat {\cal H}}_{\rm int}^{(n)}$  representing the internal state of 
$S^{(2n)}$ is 
$${\hat W}_{\rm int}^{(n)}={\otimes}_{j=1}^{(n)}
{\hat W}_{{\rm int},j},\eqno(5.1)$$
where each ${\hat W}_{{\rm int},j}$ is a copy of  the internal density matrix, 
${\hat W}_{\rm int}$ for one pair.   
\vskip 0.2cm
By Eqs. (3.5)-(3.7) and the representation by Eq. (4.5) of the positions of a pair in 
terms of its internal and external coordinates, the Hamiltonian for $S^{(2n)}$ is the 
operator ${\hat H}^{(2n)}$ in ${\hat {\cal H}}^{(2n)}$ given by the formula 
$${\hat H}^{(2n)}={\sum}_{j=1}^{n}{\hat H}_{j}^{(2)}+
{\sum}_{j,k(>j)=1}^{n}{\tilde V}(z_{j}-z_{k}{\vert}y_{j},y_{k})),\eqno(5.2)$$
where ${\hat H}_{j}^{(2)}$ is the copy of ${\hat H}^{(2)}$ representing the 
Hamiltonian of the $j$th atom and ${\tilde V}(z_{j}-z_{k}{\vert}y_{j},y_{k}))$ is 
the energy of interaction between the $j$th and $k$th atom,  given by the formula
$${\tilde V}(z_{j}-z_{k}{\vert}y_{j},y_{k})=
V\bigl(z_{j}-z_{k}+(y_{j}+y_{k})/2\bigr)+
V\bigl(z_{j}-z_{k}- (y_{j}+y_{k})/2\bigr)+$$
$$V\bigl(z_{j}-z_{k}+(y_{j}-y_{k})/2\bigr)+
 V\bigl(z_{j}-z_{k}-(y_{j}-y_{k})/2\bigr).\eqno(5.3)$$
\vskip 0.2cm 
We define the exterior Hamiltonian for $S^{(2n)}$ to be the partial mean w.r.t. the 
internal state ${\hat W}_{\rm int}^{(n)}$ of the Hamiltonian ${\hat H}^{(2n)}$, i.e.
$${\hat H}_{\rm ext}^{(n)}={\rm Tr}_{\rm int}
({\hat W}_{\rm int}^{(n)}H^{(2n)}).\eqno(5.4)$$
where ${\rm Tr}_{\rm int}$ signifies the partial trace over the Hilbert space 
${\hat {\cal H}}_{\rm int}^{(n)}$. Thus ${\hat H}_{\rm ext}^{(n)}$ is effectively 
the Hamiltonian for a system of $n$ bosonic atoms, considered as point particles. It 
follows now from Eqs. (5.1), (5.2) and (5.4), together with the normalisation 
condition for 
${\hat W}_{\rm int}^{(n)}$, that
$${\hat H}_{\rm ext}^{(n)}={\sum}_{j=1}^{n}{\hat H}_{j,{\rm ext}} 
+{\sum}_{j,k(>j)=1}^{n}{\overline V}(z_{j}-z_{k}),\eqno(5.5)$$
where ${\hat H}_{j,{\rm ext}}$ is a copy of  ${\hat H}_{\rm ext}$, as defined by 
Eq. (4.16), and  represents the exterior Hamiltonian for the $j$th atom,  and
$${\overline V}(z-z^{\prime})={\sum}_{y,y^{\prime}{\in}X}
{\rm Tr}\bigl({\hat W}_{\rm int}^{(n)}{\tilde V}(z-
z^{\prime}{\vert}y,y^{\prime})\bigr) \ {\forall} \ z,z^{\prime}{\in}X/2,\eqno(5.6)$$
It follows from Eqs. (4.16) and (5.6)  that the exterior Hamiltonian for $S^{(2n)}$ is
$${\hat H}_{\rm ext}^{(n)}=-{\hat T}{\sum}_{j=1}^{n}
{\Delta}_{{\cal V},j}+{\sum}_{j,k(>j)=1}^{n}
{\overline V}(z_{j}-z_{k}).\eqno(5.7)$$
Thus the model reduces that of a system, ${\Sigma}_{b}$, of charged  interacting 
bosons. We note that, by Eqs. (3.6) and (5.6), the interactions are of short range.
\vskip 0.5cm
\centerline {\bf 6. Conditions for ODLRO, Superconductivity and 
Antiferromagnetism.}
\vskip 0.3cm
We recall that O. Penrose and Onsager [10] presented a generalised version of Bose-
Einstein condensation, pertinent to systems of interacting bosons and subsequently 
termed off-diagonal long range order (ODLRO) by Yang [11]. For systems of 
charged particles, the combination of this property and the basic requirements of 
gauge covariance, translational invariance and thermodynamical stability implies 
superconductive electrodynamics [12-14]. The condition that a state of the charged 
bosonic system ${\Sigma}_{b}$ possess the property of ODLRO may be expressed 
in the following way. Denoting the density matrix for the state, in coordinate 
representation, by 
${\rho}(z_{1},. \ .,z_{n}{\vert}z_{1}^{\prime},. \ .,z_{n}^{\prime})$, the single 
particle density matrix is defined to be
$${\rho}_{1}(z{\vert}z^{\prime})=
{\sum}_{z_{2},. \ .,z_{n}}{\rho}(z,z_{2},. \ .,z_{n}{\vert}
z^{\prime},z_{2},. \ .,z_{n}).\eqno(6.1)$$
The ODLRO condition for the state ${\rho}$ is essentially that     
${\rho}_{1}(z{\vert}z^{\prime})$ factorises into a product of the form 
${\overline {\phi}}(z){\phi}( z^{\prime})$ when the points $z$ and $z^{\prime}$ 
are widely separated. To be precise, bearing in mind that the state ${\rho}$ depends 
on the size parameter $N$, the ODLRO condition at fixed particle density ${\eta} \ 
(:=2n/N)$ is that there exists a (non-zero) function ${\phi}$ on $X/2$ such that
$${\rm lim}_{{\vert}c{\vert}\to\infty}{\rm lim}_{N\to\infty;2n/N={\eta}}
\bigl[{\rho}_{1}(z{\vert}z^{\prime}+c)-
{\overline {\phi}}(z){\phi}(z^{\prime}+c)\bigr]=0.\eqno(6.2)$$
This condition defines ${\phi}$, which is termed the macroscopic wave function, up 
to an arbitrary constant phase factor. It has been proved to be satisfied not only by 
ideal bose gases below condensation temperatures but also by a certain class of dilute 
interacting bose gases [15, Ths. 5.1 and 7.4] in their ground states. Moreover, it has 
been cogently argued by O. Penrose and Onsager [10] that the superfluid phase of 
$He_{4}$ is characterised by ODLRO.  However, no general conditions have as yet 
been established for the occurrence of ODLRO in systems of interacting particles. In 
particular, the conditions of Lieb et al [15] are not strictly applicable to the present 
model ${\Sigma}_{b}$ in that (a) they are based on a continuum, rather than a 
lattice, model, (b) their interactions are wholly repulsive, and (c) the length, $L$, of 
the side of their containing volume is proportional to $n$ rather than $n^{1/d}$. This 
signifies that the particle density is proportional to $N^{d^{-1}-1}$, which vanishes 
in the limit $N{\rightarrow}{\infty}$. Nevertheless, we consider that (a) and (b) do 
not represent essential qualitative differences of the present model from that of Ref. 
[15] since, on the one hand, we expect the difference between the lattice and 
continuum models to be irrelevant to their large scale properties, while, on the other 
hand, the two-body interactions of ${\Sigma}_{b}$ are predominantly repulsive if 
$K/J$ is sufficiently large. However, the observation (c), as it stands, marks a serious 
difference between the two models and, in order to overcome it, we need an 
assumption to the effect that the density of ${\Sigma}_{b}$ becomes $N$-
independent in the thermodynamic limit. This appears to be quite plausible since low 
particle densities ${\eta}$ do not appear to be radically different, from the physical 
standpoint, from those of the form $N^{d^{-1}-1}$. In view of these 
observations,we make the following assumption.  
\vskip 0.3cm
(I) {\it Given that $J$ is large enough to produce Schafroth pairing in the system 
${\Sigma}$, the ground  state of  ${\Sigma}_{b}$ possesses the property of ODLRO 
if the ratio $K/J$ is sufficiently large and the density ${\eta}$  sufficiently low.}   
\vskip 0.3cm 
Further, we assume that the spin correlations of the original fermionic system 
${\Sigma}$ remain close to those of its hole-free states when the hole density 
${\eta}$ is suitably low. Thus we supplement assumption (I) with the following one. 
\vskip 0.3cm
(II) {\it At sufficiently low hole density, the ground state of the original system 
${\Sigma}$ retains its antiferromagnetic ordering.}  
\vskip 0.3cm
Thus, assuming the validity of (I) and (II), we conclude that {\it at sufficiently low 
hole density, and for a suitable range of values of the parameters $T, \ J$ and $K$, 
the system ${\Sigma}$ supports both ODLRO and magnetic ordering  at zero 
temperature.} Hence as the ODLRO property, together with the basic conditions of 
gauge covariance, translational invariance and thermodynamical stability ensures that 
the model exhibits superconductive electrodynamics [12-14], we conclude that the 
system is an antiferromagnetic superconductor at zero temperature.
 \vskip 0.5cm
\centerline {\bf 7. Concluding Remarks.}
\vskip 0.3cm
We have shown that the many electron system, ${\Sigma}$, formulated in Sections 2 
and 3, exhibits Schafroth pairing of holes; and that, consequently, the resultant pairs 
form a bosonic system, ${\Sigma}_{b}$.  Under the assumptions (I) and (II) 
discussed in Section 6, together with the conditions of sufficiently low hole density 
${\eta}$ and strong interelectronic repulsions, the ground state of ${\Sigma}_{b}$ 
possesses the property of  ODLRO, which characterises superconductivity. Hence 
assuming that, at sufficiently low hole density, the ground state of ${\Sigma}$ 
retains the magnetic ordering it possesses under the filling condition of one electron 
per lattice site, the system exhibits superconductive-cum-antiferromagnetic order.
\vskip 0.2cm
Among the challenging problems that remain are those of
\vskip 0.2cm\noindent 
(1) providing a firm basis for the assumption, made in the discussion at the end of 
Section 3, that the model $S^{(n)}$ represents the system of $n$ holes in the original 
system ${\Sigma}$; 
\vskip 0.2cm\noindent
(2) establishing the validity of assumptions (I) and (II) of Section 6 under appropriate 
conditions on the parameters of the model;
\vskip 0.2cm\noindent
(3) extending the theory to non-zero temperatures;
and
\vskip 0.2cm\noindent  
(4) providing a corresponding theory of ferromagnetic superconductivity, for which 
there is experimental evidence in Refs. [18] and [19]. 
\vskip 0.5cm
\centerline {\bf Appendix: Proof of Prop. 3.1.}
\vskip 0.3cm
As a preliminary to the proof of of Prop. 3.1, we define
$${\alpha}_{-}(x):=a(x,-s(x)) \ {\forall} \ x{\in}X\eqno(A.1)$$
and establish the following lemma.
\vskip 0.3cm
{\bf Lemma A.1.} {\it The above definitions and assumptions imply that}
$${\alpha}_{-}(x){\Phi}=0  \ {\forall} \ x{\in}X\eqno(A.2)$$  
\vskip 0.3cm
{\bf Proof.} By Eqs.(2.9), (3.1) and (A.1),
$${\alpha}_{-}(x){\alpha}^{\star}(x)+{\alpha}^{\star}(x){\alpha}_{-}(x)
={\alpha}^{\star}(x){\alpha}_{-}(x)$$
and hence
$${\alpha}_{-}(x){\alpha}^{\star}(x)=0. \ {\forall} \ x{\in}X.\eqno(A.3)$$
Furthermore, by Eq. (2.23) and (3.1), 
$${\Phi}=\bigl[{\Pi}_{x^{\prime}{\in}X}{\alpha}^{\star}(x^{\prime})\bigr]
{\Omega}$$
and, in view of the anticommutation relation (2.8) and Eq. (3.1), we may move the 
factor ${\alpha}^{\star}(x)$ of the r.h.s. of this formula to the left of all the other 
factors, where, in view of Eq. (A.3), it is annihilated by the action of 
${\alpha}_{-}(x)$. Eq. (A.2) follows immediately from this observation.
\vskip 0.3cm
{\bf Proof of Prop. 3.1.} By Eqs. (2.24) and (3.2), 
$$HC_{n}(f)=[H,D_{n}(f)]{\Phi}.\eqno(A.4)$$
where
$$D_{n}(f):=(n!)^{-1/2}{\sum}_{x_{1},. \ .,x_{n}{\in}X}f(x_{1},. \ 
.,x_{n}){\alpha}(x_{1}).\ .{\alpha}(x_{n}).\eqno(A.5))$$
Hence, defining
$$H^{(n)T}:=-T{\sum}_{j=1}^{n}{\Delta}_{{\cal V},j},\eqno(A.6)$$
$$H^{(n)J}:=-J{\sum}_{j,k(>j)=1}^{n}{\sum}_{u{\in}{\cal U}}
\bigl({\delta}(x_{j},x_{k}+u)+{\delta}(x_{j},x_{k}-u)\bigr),\eqno(A.7)$$
and
$$H^{(n)K}:=K{\sum}_{j.k(>j)=1}^{n}
{\sum}_{u,u^{\prime}({\neq}-u){\in}{\cal U}}
\bigl({\delta}(x_{j},x_{k}+u+u^{\prime})+
{\delta}(x_{j},x_{k}-u-u^{\prime})\bigr),\eqno(A.8)$$
we see from Eqs.  (2.12), (3.5) , (3.6) and (A.4)-(A.8) that in order to establish the 
formula (3.4), it suffices to prove that
$$[H^{T},D_{n}(f)]{\Phi}=D_{n}(H^{(n)T}f){\Phi},\eqno(A.9)$$
$$[H^{J},D_{n}(f)]{\Phi}=D_{n}(H^{(n)J}f){\Phi},\eqno(A.10)$$
and
$$[H^{K},D_{n}(f)]{\Phi}=D_{n}(H^{(n)K}f){\Phi},\eqno(A.11)$$
\vskip 0.3cm
{\bf Proof of Eq. (A.9).} By Eq. (A.5),
$$[H^{T},D_{n}(f)]{\Phi}=$$
$$(n!)^{-1/2}{\sum}_{x_{1},. \ .,x_{n}{\in}X}{\sum}_{j=1}^{n}
 f(x_{1},. \ .,x_{n}){\alpha}(x_{1}).. \ {\alpha}(x_{j-1})
[H^{T},{\alpha}(x_{j})]_{-}{\alpha}(x_{j+1})..{\alpha}(x_{n}){\Phi} 
\eqno(A.12)$$
and, by Eqs. (2.13) and (3.1) ,
$$[H^{T},{\alpha}(x_{j})]_{-}=
-T{\sum}_{x{\in}X,{\lambda}={\pm}1,v{\in}{\cal V}}
\bigl[a^{\star}(x,{\lambda})a(x+v,{\lambda})+a^{\star}(x+v,{\lambda})
a(x,{\lambda}),a(x_{j},s(x_{j})\bigr]_{-}$$
In view of the anticommutation relations (2.8) and (2.9), together with Eqs. (3.1) and 
(A.1), this formula reduces, after some manipulation, to the following equation.
$$[H^{T},{\alpha}(x_{j})]_{-}=
-T{\sum}_{v{\in}{\cal V}}\bigl[\bigl({\alpha}(x_{j}+v)+
{\alpha}(x_{j}-v)\bigr)
\bigl(I-{\alpha}_{-}^{\star}(x_{j}){\alpha}_{-}(x_{j})\bigr)+$$
$${\alpha}_{-}^{\star}(x_{j}){\alpha}(x_{j})\bigl({\alpha}_{-}(x_{j}+v)+
{\alpha}_{-}(x_{j}-v)\bigr)\bigr].\eqno(A.13)$$
Furthermore, by Eqs. (2.8), (3.1) and (A.1), the terms ${\alpha}_{-}(x_{j})$ and 
$({\alpha}_{-}(x_{j}+v)+{\alpha}_{-}(x_{j}-v))$ commute or anticommute with 
${\alpha}(x_{j+1}).. \ .{\alpha}(x_{n})$, according to whether $(n-j)$ is odd or 
even;  and, by Lemma (A.1), they annihilate ${\Phi}$. Consequently, by Eq. (A.13), 
the commutator $[H^{T},{\alpha}(x_{j})]_{-}$ may be replaced by 
$-T{\sum}_{v{\in}{\cal V}}\bigl({\alpha}(x_{j}+v)+{\alpha}(x_{j}-v)\bigr)$ in the 
formula (A.12). On transferring the action of this operator to the function $f$, we 
arrive at Eq. (A.9), with $H^{(n)T}$ given by Eq. (A.6), up to a harmless additive 
constsnt. 
\vskip 0.3cm
{\bf Proof of Eq. (A.10).} By Eqs. (2.14) and (A.5),
$$[H^{J},D_{n}(f)]_{-}=({n!})^{-1/2}{\sum}_{j=1}^{n}
f(x_{1},. \ .,x_{n}){\alpha}(x_{1}). \ .{\alpha}(x_{j-1})
[H^{J},{\alpha}(x_{j})]_{-}{\alpha}(x_{j+1}). \ .{\alpha}(x_{n})
\eqno(A.14)$$
and, by Eqs. (2.7)-(2.9),  (2.14) and (3.1),
$$[H^{J},{\alpha}(x_{j})]_{-}=J{\sum}_{x{\in}X,u{\in}
{\cal U}}\bigl({\sigma}(x)[{\sigma}(x+u),{\alpha}(x_{j})]_{-}+
[{\sigma}(x),{\alpha}(x_{j})]_{-}{\sigma}(x+u)\bigr)=$$
$$Js(x_{j}){\sum}_{x{\in}X,u{\in}
{\cal U}}\bigl({\delta}(x+u,x_{j}){\sigma}(x)
{\alpha}(x_{j})+{\delta}(x,x_{j}){\alpha}(x_{j}){\sigma}(x+u)\bigr)=$$
$$Js(x_{j}){\sum}_{u{\in}{\cal U}}\bigl({\sigma}(x_{j}-u){\alpha}(x_{j})+
{\alpha}(x_{j}){\sigma}(x_{j}+u)\bigr).$$
Hence since, by Eqs. (2.5) , (2.7)-(2.9) and (3.1), ${\alpha}(x_{j})$ commutes with 
${\sigma}(x_{j}-u)$,  
$$[H^{J},{\alpha}(x_{j})]_{-}=Js(x_{j})
{\sum}_{u{\in}{\cal U}}{\alpha}(x_{j})\bigl({\sigma}(x-u)+{\sigma}(x+u)\bigr).
\eqno(A.15)$$
Further, since
$$[{\sigma}(x_{j}{\pm}u),{\alpha}(x_{j+1}).. \ .{\alpha}(x_{n})]_{-}{\Phi}=$$
$${\sum}_{k=j+1}^{n}{\alpha}(x_{j+1}). \ .,{\alpha}(x_{k})
[{\sigma}(x_{j}{\pm}u),{\alpha}(x_{k})]_{-}{\alpha}(x_{k+1}). \ 
.{\alpha}(x_{n}){\Phi}\eqno(A.16)$$
and since, by Eqs. (2.5), (2.7)-(2.9) and (3.1), 
$$[{\sigma}(x_{j}{\pm}u),{\alpha}(x_{k})]_{-}=
{\delta}(x_{j}{\pm}u,x_{k})s(x_{j}{\pm}u){\alpha}(x_{k}),\eqno(A.17)$$
 it follows from Eqs. (2.21), (A.16) and (A.17) that 
$${\sigma}(x_{j}{\pm}u){\alpha}(x_{j+1}). \ .{\alpha}(x_{n}){\Phi}=$$
$$s(x_{j}{\pm}u){\alpha}(x_{j+1}). \ .{\alpha}(x_{n}){\Phi}-
 {\sum}_{k=j+1}^{n}{\delta}(x_{j}{\pm}u,x_{k})s(x_{j}{\pm}u)
{\alpha}(x_{j+1}). \ .{\alpha}(x_{n}){\Phi}.\eqno(A.18)$$
Since, for $u{\in}{\cal U}$, the sites $x_{j}$ and $(x_{j}+u)$ are nearest 
neighbours and therefore carry opposite spins in the antiferromagnetic configuration, 
$$s(x_{j})s(x_{j}-u)=-1.\eqno(A.19)$$. 
Hence, by Eqs. (A.15), (A.18) and (A.19), 
$$[H^{J},{\alpha}(x_{j})]_{-}{\alpha}(x_{j+1}). \ .{\alpha}(x_{n}){\Phi}=$$
$$4dJ{\alpha}(x_{j}). \ .{\alpha}(x_{n}){\Phi}-
J{\sum}_{u{\in}{\cal U}}{\sum}_{k=j+1}^{n}\bigl({\delta}(x_{j},x_{k}+u)+
{\delta}(x_{j},x_{k}-u)\bigr){\alpha}(x_{j}). \ .{\alpha}(x_{n}){\Phi}$$
It follows from this equation and Eq. (A.14) that Eq. (A.10) is satisfied, with 
$H^{(n)J}$ given by Eq. (A.4), up to an irrelevant additive constant $4dJ$. 
\vskip 0.3cm
{\bf Proof of Eq. (A.11).} Noting that, by Eqs. (2.5), (2.6) and (3.1),
$$[{\nu}(x),{\alpha}(x^{\prime})]_{-}=-{\alpha}(x){\delta}(x,x^{\prime}),
\eqno(A.20)$$
we see that it is a simple matter to prove (A.11) by replacing $s(x), \ {\sigma}(x), \ 
u$ and $J$ by ${\nu}, \ 1, \ u+u^{\prime}$ and $K$, respectively, in the proof of 
(A.10). This completes the proof of the Proposition.
\vskip 0.5cm
{\bf Acknowledgement.} The author wishes to thank Walter Wreszinski for some 
helpful remarks about an earlier draft of this article 
\vskip 0.3cm
\centerline {\bf References}
\vskip 0.3cm\noindent
[1] J.G. Bednorz and K. A. Mueller: Z. Phys. B {\bf 64}, 189. 1986
\vskip 0.2cm\noindent
[2] P. W. Anderson: {\it The theory of superconductivity in the high $T_{c}$ 
cuprates}, Princeton Univ. Press, 1997
\vskip 0.2cm\noindent
[3] N. M. Plakida: {\it High temperature superconductivity}, Springrer, Berlin, 1994
\vskip 0.2cm\noindent
[4] G. Baskaran, Z. Zou and P. W. Anderson: Solid State Comms. {\bf 63}, 973, 
1967
\vskip 0.2cm\noindent
[5] S. A. Kivelson, D. S. Rokhsar and J. P. Sethna: Phys. Rev. B {\bf 35}, 8865, 
1987
\vskip 0.2cm\noindent
[6] P. Richmond and G. L. Sewell: J. Math. Phys. {\bf 9}, 349, 1968
\vskip 0.2cm\noindent
[7] M. R. Schafroth: Phys. Rev. {\bf 100}, 463, 1955
\vskip 0.2cm\noindent
[8] L. N. Cooper: Phys. Rev. {\bf 104}, 1189, 1956
\vskip 0.2cm\noindent
[9] A. J. Leggett: Lec. Notes in Physics, Vol. 115, Pp. 13-27, 1980
\vskip 0.2cm\noindent
[10] O. Penrose and L. Onsager: Phys. Rev. {\bf 104}, 576, 1956
\vskip 0.2cm\noindent
[11]  C. N. Yang: Rev. Mod. Phys. {\bf 34},694, 1962 
\vskip 0.2cm\noindent
[12] G. L. Sewell: J. Stat. Phys. {\bf 61}, 415, 1990
\vskip 0.2cm\noindent
[13] G. L. Sewell: J. Math. Phys. {\bf 38}, 2053, 1997
\vskip 0.2cm\noindent
[14] G. L. Sewell: {\it Quantum mechanics and its emergent macrophysics}, 
Princeton Univ. Press, Princeton, 2003
\vskip 0.2cm\noindent
 [15] E. H. Lieb, R. Seiringer, J. P. Solovej and J. Yngvason: {\it The mathematics of 
the bose gas and its condensation}, Birkhauser, Basel, 2005
\vskip 0.2cm\noindent
[16] S. A. Kivelson, G. Aeppli and V. J. Emery:Proc. Nat. Acad. Sci. {\bf 98},  
11903, 2001
\vskip 0.2cm\noindent
 [17] M. Gutzwiller: Phys. Rev. {\bf 137}, A1726, 1965
\vskip 0.2cm\noindent
[18] S. S. Saxena, P. Agarwal, K. Ahilan, F. M. Grosche, R. K. W. Haselwimmer, M. 
J. Steiner, E. Pugh, I. R. Walker, S. R. Julian, P. Monthoux, G. G. Lonzarich, A. 
Huxley, I. Shelkin, D. Braithwaite and J. Flouquet: Nature {\bf 406}, 587, 2000
\vskip 0.2cm\noindent
[19] D. Aoki and J. Flouquet: J. Phys. Soc. Japan{\bf 81}, 011003, 2012

\end